\documentclass[english,english,noshowpacs, amssymb,twocolumn,aps,nofootinbib]{revtex4}

\usepackage[T1]{fontenc}
\usepackage[latin9]{inputenc}
\usepackage{color}
\usepackage{mathtools}
\usepackage{textcomp}
\usepackage{amsmath}
\usepackage{esint}
\usepackage{url}
\usepackage{natbib}
\usepackage{graphicx}
\usepackage[export]{adjustbox}

\begin{document}

\title{Atomic Resonant Single-Mode Squeezed Light from \\ Four-Wave Mixing through Feedforward}

\author{Saesun Kim and Alberto M. Marino}

\affiliation{Homer L. Dodge Department of Physics and Astronomy, The University of Oklahoma, Norman, Oklahoma 73019, USA}

\begin{abstract}
Squeezed states of light have received renewed attention due to their applicability to quantum-enhanced sensing. To take full advantage of their reduced noise properties to enhance atomic-based sensors, it is necessary to generate narrowband near or on atomic resonance single-mode squeezed states of light. We have previously generated bright two-mode squeezed states of light, or twin beams, that can be tuned to resonance with the D1 line of $^{87}$Rb with a non-degenerate four-wave mixing (FWM) process in a double-lambda configuration in a $^{85}$Rb vapor cell. Here we report on the use of feedforward to transfer the amplitude quantum correlations present in the twin beams to a single beam for the generation of single-mode amplitude squeezed light. With this technique we obtain a single-mode squeezed state with a squeezing level of $-2.9\pm0.1$~dB when it is tuned off-resonance and a level of $-2.0\pm 0.1$~dB when it is tuned on resonance with the D1 $F=2$ to $F'=2$ transition of $^{87}$Rb.
\end{abstract}

\maketitle

\section{Introduction}

Squeezed states of light have recently received considerable attention due to their applicability to quantum metrology given their reduced noise properties.  For example, squeezed light has been used to enhance the sensitivity of gravitational wave measurements~\cite{LIGO} and as the basis for applications such as quantum computation~\cite{PfisterCom}, quantum communications~\cite{SchnabelKey}, and quantum imaging~\cite{MarinoImg}. Strong coupling between squeezed light and an atomic system is needed to extend its applicability to atomic magnetometers~\cite{MitchellSen}, atomic interferometers~\cite{HaineAtomIn}, and spectroscopy~\cite{KimbleSpec}. The ability to implement a strong coupling will also enables the possibility to explore phenomena such as non-classical excitation of atoms~\cite{ParkinsNoncl}, entanglement transfer~\cite{KimEnt}, and the interaction with an atomic system in an electromagnetically induced transparency (EIT) configuration~\cite{BienertEIT}.

In order to have a strong coupling between squeezed light and an atomic system, it is necessary to generate narrowband squeezed light on atomic resonance. However, its generation with an optical parametric oscillator (OPO)~\cite{GiacobinoOPO,SchnabelOPO}, the most common technique for the generation of squeezed light, is not efficient due to the losses in the nonlinear crystal of the required UV pump light~\cite{LamResOPO,MitchellResOPO}. In addition, it is necessary to engineer the nonlinear crystal to operate at the desired wavelength and obtain efficient narrowband operation. Despite these challenges, narrowband on-resonance squeezed light has been generated with an OPO. For example, single-mode vacuum squeezed light resonant with the Rb D1 line ($-5.6$~dB of squeezing) and Cs D2 line ($-3$~dB of squeezing) has been generated~\cite{LamResOPO,WangResOPO,GiacobinoResOPO}.

An alternative to the use of an OPO to generate squeezed light is the use of an atomic ensemble.  In principle this is an attractive option as it naturally generates narrowband resonant squeezed light. However, due to the strong absorption associated with a resonant transition, losses are a limiting factor for the generation of resonant squeezed light~\cite{NovikovaPSR}. Nevertheless, resonant vacuum single-mode squeezing has been reported using polarization self-rotation (PSR) in a Rb vapor cell. Although theoretical predictions suggest PSR can generate $-6$~dB of squeezing~\cite{RochesterPSR}, only $-2.9$~dB of squeezing on resonance with the D1 lines of Rb has been generated~\cite{LezamaPSR}. A recent theoretical analysis shows that noise from the thermal vapor can limit the levels of squeezing that can be achieved with PSR to levels significantly below original predictions~\cite{NussenzveigPSR}.

\begin{figure}
  \centering
  \includegraphics[width=\columnwidth]{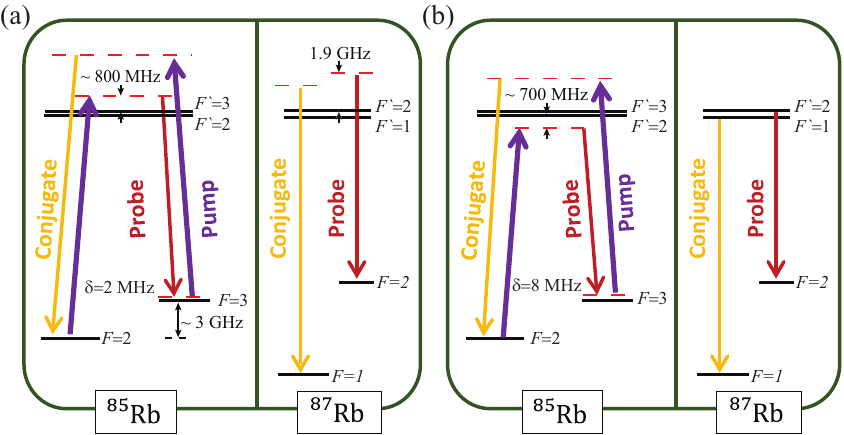}
  \caption{Energy level diagram of the double-lambda system in the D1 line of $^{85}$Rb. Optimized FWM configuration for the generation of (a) off-resonance two-mode squeezed states with $-7.4\pm0.1$~dB of squeezing and (b) two-mode squeezed states with $-5.8\pm0.1$~dB of squeezing with a probe resonant with the D1 $F=2$ to $F'=2$ transition in the D1 line of $^{87}$Rb.}
  \label{figure1}
\end{figure}

While only moderate levels of single-mode squeezing have been generated with an atomic ensemble, off-resonant two-mode states of light with up to $-9$~dB of intensity-difference squeezing have been generated using a non-degenerate four-wave mixing (FWM) process in a double-lambda configuration in $^{85}$Rb and $^{87}$Rb~\cite{LettFWM1}. By taking advantage of the proximity of the energy levels of the two rubidium isotopes, we have recently shown that it is possible to operate the FWM in $^{85}$Rb in a regime in which one of the modes can be resonant with the $F=2$ to $F'=2$ or $F=1$ to $F'=1$ transition in the D1 line of $^{87}$Rb, as shown in Fig.~\ref{figure1}, with $-6.3$~dB and $-6.2$~dB of intensity-difference squeezing after subtracting electronic noise, respectively ~\cite{MarinoFWM}. Although there is significant interest in generating two-mode squeezed light due to its entanglement property, it has been shown that single-mode squeezing is in general better for sensing applications~\cite{FabreUltimate}.

Given the presence of quantum correlations between the amplitudes and phases of the twin beams, it is possible to transfer the squeezing form a two-mode configuration to a one-mode configuration by measuring one of the beams and using the information to perform a feedforward operation on the other.  In particular, if we consider the amplitude correlations present in the twin beams, we have that when one mode has a large positive fluctuation, the other mode will have a corresponding fluctuation.  Thus if we measure the amplitude fluctuations of one of the modes, we can use that information to decrease the fluctuations in the other through feedforward. Due to the quantum nature of the correlations, it is possible to generate a single-mode squeezed beam with this technique~\cite{ReynaudFeed,FabreFeed}.

One of the advantages of the feedforward technique is that it is possible to leverage the large levels of two-mode squeezing that have been generated with atomic systems for the generation of single-mode squeezing. In the ideal case, the absolute noise of the single beam after the feedforward can be reduced to the absolute noise level of the initial twin beams. However, due to the fact that the beam after feedforward has about half the power as the initial twin beams, the level of squeezing will decrease by $3$~dB. Thus, given that two-mode squeezed light with over $-9$~dB of squeezing has been generated with an atomic system with FWM, feedforward
can theoretically generate over $-6$~dB of narrowband single-mode squeezing even with the $3$~dB penalty. In addition to the promise of higher squeezing levels, squeezed light from feedforward can generate bright quantum states of light~\cite{GaoFeed}, which provide an advantage for sensing and applications that are based on intensity measurements~\cite{XiaoSqueezing}.

In this letter, we generate single-mode amplitude squeezed light through feedforward. We obtain a squeezing level of $-2.9\pm0.1$~dB when the squeezed light is tuned around 1.9~GHz away from the $F'=2$ to $F=2$ transition in the D1 line of $^{87}$Rb and a squeezing level of $-2.0\pm0.1$~dB when it is on resonance with this transition.

\section{Experimental Results}

We generate bright two-mode squeezed states of light using a non-degenerate FWM process in a double-lambda configuration in the D1 line of $^{85}$Rb~\cite{LettFWM2}. In this configuration, two photons from a single pump beam are converted into two photons, one for the probe and one for the conjugate. As a result, the input probe is amplified and a new field, the conjugate, is generated.  The simultaneous emission of probe and conjugate photons leads to quantum correlations between them.

\begin{figure}
\centering
\includegraphics[width=\columnwidth]{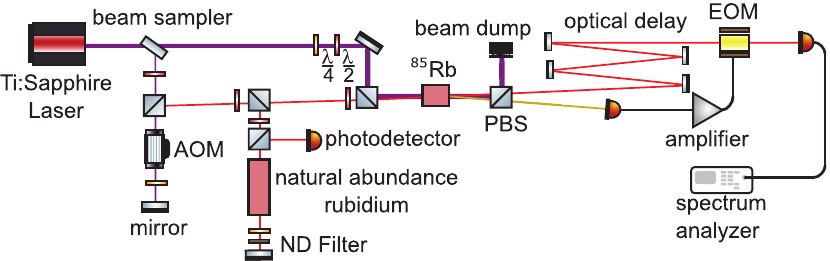}
\caption{Simplified feedforward experimental setup. A FWM process is used to generate quantum correlated probe and conjugate beams. After the FWM, the conjugate is detected with a photodiode and information on the conjugate amplitude fluctuations is then used to control the amplitude of the probe with an electro-optic modulator (EOM) to implement the feedforward.  An optical delay line in the path of the probe is used to compensate the electronic delay. A saturated absorption spectroscopy setup is used to obtain an absolute frequency measure for the probe. AOM: acousto-optic modulator; ND filter: neutral density filter; PBS: polarizing beam splitter.}
\label{figure2}
\end{figure}

As shown in Fig.~\ref{figure2}, we use a CW Ti:Sapphire laser to generate the pump beam for the FWM. A small portion of the laser is picked off with a beam sampler and double passed through an acoustic optical modulator (AOM) to generate the probe with the required frequency. The frequency shift is set such that the probe and pump are effectively on two-photon resonance between the two ground state hyperfine levels of $^{85}$Rb.  Due to the light shift introduced by the pump, this translates to a two-photon detuing $\delta$ from the bare atomic levels of a few MHz.

Both the probe and pump are sent through optical fibers to clean up their spacial modes before they are made to intersect at an angle $\theta$ at the center of a 12~mm $^{85}$Rb vapor cell with anti-reflection coated windows to implement the FWM. After the FWM process, the pump is blocked with a polarization filter and the conjugate field is detected with a 95\% quantum efficient photodetector to measure its amplitude fluctuations. The probe beam is sent through an electro-optic modulator (EOM) to feedforward the conjugate fluctuations onto its amplitude.  To perform the feedforward operation, we block the DC component of the conjugate photocurrent such that only the information of the conjugate amplitude fluctuations remain.  The resulting signal is then amplified with a variable gain and used to control the EOM transmission.  Since the amplitude fluctuations of the probe and conjugate are quantum correlated, the noise level of the probe beam after the feedforwad will be below the quantum noise limit (QNL). Reaching this level of reduced noise requires proper optimization of the feedforward gain.

We consider two different configurations for the FWM source to obtain on resonance and off-resonance single-mode squeezing after the feedforward. In order to obtain an absolute measure or the probe frequency, we take a portion of the probe beam before the FWM process and send it to a saturated absorption spectroscopy setup, as shown in Fig.~\ref{figure2}.

\subsection{Off-Resonance Configuration}

We first consider the off-resonance configuration in which the probe beam is detuned from the D1 $F=2$ to $F'=2$ transition of $^{87}$Rb by around 1.9~GHz, as shown in Fig.~\ref{figure1}(a). In this configuration the FWM operates at a one-photon detuning of around 800~MHz to the blue of the D1 line of $^{85}$Rb and a two-photon detuning $\delta$ of 2~MHz. An intensity-difference squeezing of $-7.4\pm0.1$~dB (after electronic noise subtraction) at an analysis frequency of 360~kHz is obtained for an angle $\theta$ of 0.5$^{\circ}$, pump power of 300~mW, vapor cell temperature of 107$^{\circ}$C, and $1/e^2$ pump and probe diameters of 0.9~mm and 1.3~mm, respectively.  

After generation of the twin beams we perform the feedforward as described above. Given that the electronics used for the feedforward have a finite time response, there is a time delay introduced on the signal for the conjugate fluctuations. A delay in the time domain translates to a phase shift  in the frequency domain that effectively makes the sign of the feedforward gain oscillate as a function of frequency. This would result in a probe noise level after the feedforward that oscillates between the absolute probe-conjugate intensity-difference and intensity-sum noise levels. Thus, if the electronic delay is not compensated, the noise spectrum of the probe after feedfoward will oscillate between squeezing and excess noise. To compensate for this delay, we introduce an optical delay line on the probe before sending it through the EOM.  We optimize the optical path to minimize the oscillations and find that a propagation length of 19.6~m is needed, which represents a delay of $\approx65$~ns.

The noise spectrum of the probe after feedforward can be modeled with a semiclassical formalism~\cite{FabreFeed}. If we assume the optimum feedforward gain at every analysis frequency $\Omega$, $G_{opt}(\Omega )=\sqrt{\eta_E}(S_{-}(\Omega )-S(\Omega ))/[\eta_D~S(\Omega )+(1-\eta_D)]$, and that the probe and conjugate normalized noise spectra are equal (as expected for the ideal case), then the normalized probe noise spectrum after feedforward $S_f$ is given by
\begin{equation}
S_f(\Omega )=\eta_E~S(\Omega )+(1-\eta_E)-\eta_E~\eta_D~\frac{(S(\Omega )-S_{-}(\Omega ))^2}{\eta_D~S(\Omega )+(1-\eta_D)},
\label{prediction}
\end{equation}
\noindent
where $S(\Omega )$ and $S_{-}(\Omega )$ are the single beam intensity noise and the twin beam intensity-difference noise spectra, respectively. All the noise spectra in this equation are normalized to their corresponding shot noise. The first two terms on the right-hand side of Eq.~(\ref{prediction}) correspond to the contribution of the vacuum fluctuations that couple in due to the optical losses in the path of the probe from the delay line and EOM ($\eta_E=0.88$), while the last term represents the noise cancellation due to the feedforward when the conjugate is measured with a detector with efficiency $\eta_D=0.95$.

Since the normalized single-beam noise is significantly larger than the QNL ($S(\Omega ) \gg 1$) in the limit of high FWM gain needed to obtain large levels of squeezing, in the ideal case of no losses ($\eta_E=\eta_D=1$)
\begin{align}
S_{f}(\Omega )=2 S_{-}(\Omega ),
\label{prediction2}
\end{align}
which represents the 3~dB penalty in transferring the squeezing from a two-mode to a single-mode configuration. This penalty limits the maximum amount of single-mode squeezing we can obtain from the feedforward technique.

Figure~\ref{4-2feed}(a) shows the measured intensity noise spectra normalized to the corresponding shot noise after electronic noise subtraction. As expected, the normalized single beam noise $S(\Omega )$ (purple trace) is significantly higher than the QNL (green trace) given the large gain of the FWM process. As can be seen, single-mode squeezing is obtained after feedforward both with (black trace) and without (grey trace) the optical delay line. The peaks around 1~MHz are due to a piezo-electric resonances of the EOM used for the feedforward. It is worth nothing that without optical delay single-mode squeezing can only be obtained in the low-frequency region given that the noise increases as a function of frequency due to the phase shift introduced by the electronic delay. Given the 65~ns delay we need to compensate for, a full oscillation of the noise spectrum would happen at a frequency of 15.3~MHz, so only the initial portion of that oscillation can be seen in Fig.~\ref{4-2feed}(a). This limits the measured level of single-mode squeezing without the optical delay line to $-0.8$~dB at an analysis frequency of 360~kHz. When we introduce the optical delay line the level of single-mode squeezed is increased to $-2.9\pm0.1$~dB.  Additionally, squeezing can now be observed from at least 200~kHz to over 2.2~MHz.

We can obtain the theoretical prediction for the probe noise spectrum after feedforward (red trace) by using the measure single beam noise spectra $S(\Omega )$ (purple trace) and the intensity-difference squeezing $S_{-}(\Omega )$ (blue trace) in combination with the measured losses and Eq.~(\ref{prediction}).  As can be seen, the noise level of the optimized measured single-mode squeezing spectrum increases faster with frequency than expected from the theoretical prediction.  This illustrates one of the complications of the feedforward technique; that is, that the feedforward gain has to be optimized for every frequency to obtain the best squeezing level and largest bandwidth.  In our experiment the bandwidth of the detector used to measure the conjugate is  $\sim4$~MHz, so effectively the feedforward gain decreases as the analysis frequency approaches the detector bandwidth.

As can be seen from the predicted noise spectrum for the probe after feedforward, the expected level of squeezing of $-3.5$~dB at an analysis frequency of 360~kHz represents a reduction of more than $3$~dB from the initial level of intensity-difference squeezing of $-7.4\pm0.1$~dB, as given by Eq.~(\ref{prediction2}).  The additional reduction in the level of squeezing is a result of the optical losses in the feedforward and the quantum efficiency of the probe detector. In principle it is possible to minimize these losses by implementing the feedforwad through a displacement operation by combining the probe beam with a strong coherent state on a 99/1 beamspliter~\cite{GaoFeed}.  In this case the expected level of squeezing would increase to $-4.3$~dB.

\subsection{On-Resonance Configuration}

For the on-resonance configuration, shown in Fig.~\ref{figure1}(b), the one-photon detuning of the FWM is set $687$~MHz to the red of the D1 $F=2$ to $F'=2$ transition of $^{85}$Rb in order for the probe to be resonant with the D1 $F=2$ to $F'=2$ transition in $^{87}$Rb.
We obtain the optimum level of squeezing for a two-photon detuning $\delta$ of 8~MHz (which compensates for the light shift in this configration as shown in our previous work~\cite{MarinoFWM}), an angle $\theta$ of 0.4$^{\circ}$, pump power and diameter of 750~mW and 1.3~mm, respectively, probe diameter of 0.9~mm, and vapor cell temperature of 90$^{\circ}$C. With these parameters we obtain an intensity-difference squeezing level of $-5.8\pm0.1$~dB at an analysis frequency of 360~kHz after subtracting the electronic noise.

\begin{figure}[t!]
  \centering
  \includegraphics[width=\columnwidth]{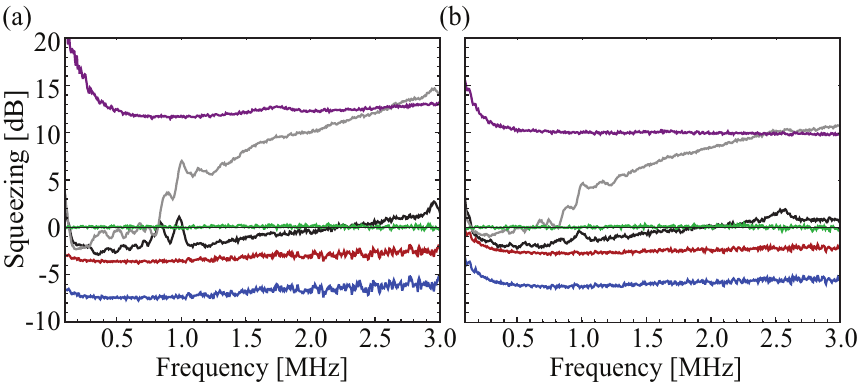}
  \caption{Normalized intensity noise spectra for the (a) off-resonance and (b) on-resonance configurations.  Each trace represents: conjugate noise (purple trace), quantum noise limit (green trace), probe noise after feedforward with optical delay (black trace) and without optical delay (gray trace), twin beam intensity-difference squeezing (blue trace), and prediction for probe noise after feedforward (red trace). The electronic noise has been subtracted from all trances and the spectrum analyzer was set to average 10 traces with a resolution bandwidth of 30~kHz and a video bandwidth of 100~Hz.}
  \label{4-2feed}
\end{figure}

For this configuration, the optimum length of the optical delay line was found to be 21.7~m, which represents a delay of $\approx72$~ns. The slightly difference in the delay between the off-resonance an on-resonance configurations is due to the fact that a portion of the delay results from the FWM process itself. As shown in~\cite{LettFWM3}, for the off-resonance configuration the FWM introduces a relative delay of the probe with respect to the conjugate of $\sim10$~ns, consistent with a $\approx7$~ns delay we need to compensate for to obtain the optimum intensity-difference squeezing for this configuration. On the other hand, we find that no delay is needed to obtain the optimum intensity-difference squeezing for the on-resonance configuration for the FWM.

Figure~\ref{4-2feed}(b) shows the normalized measured intensity noise spectra for the resonant squeezed light configuration. The measured level of squeezing of $-0.5$~dB at an analysis frequency of 360~kHz after feedforward without a delay line increases to $-2.0\pm0.1$~dB when the delay line is introduced. For this configuration, the theoretical prediction for the level of single-mode squeezing after feedforward is of $-2.4$~dB. This level increases to $-2.8$~dB if losses can be minimized through the displacement operation technique to implement the feedfowrard.
Similar to the off-resonant configuration, without the optical delay line we can only obtain single-mode squeezing in the low-frequency region as the noise increases rapidly with frequency. The delay line allows us to observe squeezing from at least 200~kHz to over 2.0~MHz, which is a slightly small range than for the off-resonant configuration due to the smaller amount of initial squeezing.

One of the advantages of the system we have implemented is that it allows some tunability around resonance. Previous papers show that it is possible to obtain off-resonance two-mode squeezing over a region of more than 1~GHz~\cite{LettFWM1} and on-resonance two-mode squeezing over a region of the order of 500~MHz~\cite{MarinoFWM}. The tunability of the two-mode squeezed light can be transferred to the single-mode squeezed light. Additionally, if we were to measure the probe and feedforwad on the conjugate we should be able to generate single-mode squeezed light on resonance with the D1 $F=1$ to $F'=1$ transition in $^{87}$Rb.

\section{Conclusion}
We have demonstrated the generation of bright single-mode squeezed light using a non-degenerate FWM process in the D1 line of $^{85}$Rb in combination with a feedforward technique. For the resonant configuration in which the probe is on resonance with the $F=2$ to $F'=2$ transition in the D1 line of $^{87}$Rb, we obtain $-2.0\pm0.1$~dB of single-mode squeezing. The squeezing level increases to $-2.9\pm0.1$~dB when we tuned the source off resonance. The obtained levels of squeezing are limited by losses in the implementation of the feedforward.  Minimizing these losses would in theory allow us to obtain squeezing levels of $-2.8$~dB for the on-resonance configuration and $-4.3$~dB  for the off-resonance configuration. The combination of feedforward and the atomic based FWM process can in principle offer large levels of narrowband squeezing on atomic resonance while overcoming several of the experimental difficulties associated with crystal-based sources. Resonant squeezed states with bandwidths of the order of the atomic natural linewidth will allow for an efficient and strong interaction between atoms and quantum states of light that will make it possible to enhance the sensitivity of the atomic-based sensors.

\acknowledgments

This work was supported by a grant from the AFOSR (FA9550-15-1-0402). The authors would like to thank Ashok Kumar for constructive criticism of the manuscript and Tim Woodworth and Gaurav Nirala for insightful conversations.

\end{document}